\documentclass[graybox]{svmult}

\usepackage{mathptmx}       
\usepackage{helvet}         
\usepackage{courier}        
\usepackage{type1cm}        
\usepackage{makeidx}         
\usepackage{graphicx}        
\usepackage{multicol}        
\usepackage[bottom]{footmisc}

\makeindex             

\begin{document}

\title*{The Nature of Faint Blue Stars in the PHL and Ton Catalogues based on Digital Sky Surveys}
\author{H.\ Andernach, F.\ Romero Sauri, W.\ Cop\'o C\'ordova, I.\ del C.\ Santiago-Bautista}
\authorrunning{The Nature of Faint Blue Stars in the PHL/Ton Catalogues} 
\institute{Heinz Andernach \at Deptartamento de Astronom\'{i}a, DCNE, Universidad de Guanajuato, Guanajuato, Mexico \\
  \email{heinz@astro.ugto.mx}
\and Felipe Romero Sauri \at Universidad Aut\'onoma de Yucat\'an, M\'erida, Mexico  
\and Wilbert Cop\'o C\'ordova \at Instituto Tecnol\'ogico Superior de Centla, Tabasco, Mexico  
\and Iris del Carmen Santiago-Bautista \at Departamento de Astronom\'{i}a, DCNE, Universidad de Guanajuato, Guanajuato, Mexico
}
%
%
\maketitle

\abstract{We determined accurate positions for 3000 of the ``faint blue stars'' 
in the PHL (Palomar-Haro-Luyten) and Ton/Ton\,S catalogues. These were 
published from 1957 to 1962, and, aimed at finding new white dwarfs, provide
approximate positions for $\sim$10,750 blue stellar objects. Some of
these ``stars'' had become known as quasars, a type of objects unheard-of
before 1963. We derived subarcsec positions from a comparison of published
finding charts with images from the first-epoch Digitized Sky Survey.
Numerous objects are now well known, but unfortunately neither their PHL 
or Ton numbers, nor their discoverers, are recognized in current databases.
A comparison with modern radio, IR, UV and X-ray surveys leads us to suggest
that the fraction of extragalactic objects in the PHL and Ton catalogues is
at least 15\,\%. However, because we failed to locate the original PHL
plates or finding charts, it may be impossible to correctly identify the 
remaining 7726 PHL objects.  }

\section{Introduction and Motivation for this Work}
\label{sec:1}

In 1947 Humason \& Zwicky \cite{hz47} searched for faint blue stars
using 4-filter photography, with the aim to find new white dwarfs
(WDs). Motivated by this, Luyten \cite{luyten53} surveyed the north
Galactic cap for faint blue stars with red and blue plates.  This, in
turn, prompted G. Haro to search systematically for blue stars using
three exposures on the same plate. This resulted in (a) the
``Ton'' and ``Ton\,S'' lists of 2008 faint blue objects in the north
\cite{irichav57,chav59} and south \cite{chav58} Galactic caps from
Tonantzintla Schmidt telescope plates, and (b) a list of 8746 ``PHL'' objects
from 49 Palomar 48-inch Schmidt plates \cite{hl62}.  One year later, the
first radio-loud quasar was reported \cite{schmidt63}, and within another
two years, radio-quiet quasars were found to be $\sim$10 times more
common than radio-loud ones \cite{sandage65}.  Some of the PHL or Ton
objects became very famous, like, e.g., the high-redshift QSOs PHL\,2871
(3C\,9) or PHL\,957. In fact, 25 PHL objects have over 200 references
in Simbad \cite{simbad} ($\sim$50\,\% are Seyfert galaxies or QSOs, and
the remainder are Galactic stars).

While Simbad contains all 8746 PHL objects, 94.6\,\% of these are listed
with their poor original position ($\pm\sim1.5'$), and only 490 (5.6\%)
appear with more precise positions (150 QSOs or AGN, 160 WDs, and 180 
stars of other types). On the other hand, NED \cite{ned} recognizes only 302
(3.5\,\%) of all PHL objects (113 QSOs, 21 galaxies, 144 stars, 8 WDs and
16 of unknown type). Unfortunately, neither NED nor Simbad quote the detection
paper \cite{hl62} for {\it any} of the PHL/Ton objects they contain. This
motivated us to try to determine precise positions for those PHL and
Ton objects with published finding charts (FCs) in order to: (a) assess
how many are already known (with other names) in astronomical databases,
(b) estimate the fraction of extragalactic objects among them, and (c)
give credit to the original discoverers.

\section{Procedure and Results}
\label{sec:2}

While the Ton and Ton\,S catalogues \cite{irichav57,chav59,chav58} included
FCs for all 2008 objects, only for 1020 (12\,\%) of all PHL objects FCs were
published \cite{hc87,chav88,chav90,chav92}. To identify the correct PHL or Ton
object, we displayed the published FCs side-by-side to a similar-sized 
(16$'\times16'$) image of the Digitized Sky Survey (DSS) centered on the 
published position, and then retrieved the position of the marked object
to $<1''$ precision. We chose the blue, first-epoch ``DSS1'', for being most 
similar to the published FCs in both color and epoch, to avoid displacements for 
high proper motion objects.

We were able to identify all 1020 PHL objects, but, owing to some very
poor FCs, only $\sim$97\,\% of the Ton/TonS objects; a few objects were
found with the help of astrometry.net \cite{lang10}. The mean positional
offsets for the PHL and Ton/Ton\,S objects (``published minus DSS'') 
are 21$''$ and $7''$ in R.A., and $4''$ and $-11''$ in Decl., with a 
dispersion in position of $\sim1'$ for PHL and $\sim2'$ for Ton/Ton\,S.

During our work we found several curiosities, like {\it very large
positional offsets}, often due to typos which can be resolved using
\cite{lang10}; {\it sign errors}: e.g., the Decl. sign for
PHL\,1 should be negative, since only with this choice there is a blue
object near to its published position; {\it large proper motions} of
$\ge3''$ between the epochs of DSS1 ($\sim$1950) and SDSS ($\sim$2000;
\cite{ahn2012}) implying proper motions in excess of 60\,mas\,yr$^{-1}$
for several objects; {\it variability}: a few objects appear on some
plates, but not on others, e.g., PHL\,6287 is present on DSS1 images, but
absent in DSS2 and SDSS.  We found some of the cross-identifications
with PHL objects proposed by \cite{bf84} to be false, as can be seen on
FCs in \cite{hc87,chav88,chav90,chav92}, which were published {\it after}
\cite{bf84}.

\section{Conclusions}
\label{sec:3}

For the first time since their publication, we derived accurate positions 
for all 3000 PHL/Ton objects with available finding charts (FCs).
We found erroneous identifications for PHL objects in the literature,
caused by the lack of an available FC. Thus, the latter are crucial for 
an unambiguous identification. Many PHL/Ton objects have become 
well-known objects, but for none of them the 
discovery papers are cited in NED or Simbad.
Curiously, the best-known PHL objects are {\it not} those with published FCs
which suggests that their positions had either been found from the approximate 
original ones, or that FCs had been circulated privately by \cite{hl62}.
No traces could be found of the 49 PHL plates, neither
at INAOE (Tonantzintla, Mexico, where Haro and Chavira worked), nor at Univ.\ of Minnesota 
(where Luyten worked), nor of the corresponding FCs for the remaining 
7726 (88\,\% of all) PHL objects This makes the correct identification 
of these objects virtually impossible, except perhaps for the $\sim$1400 
``very blue'' ones in table~2 of \cite{hl62}.

Extragalactic objects may be distinguished from Galactic stars using 
color indices from IR to UV wavelengths (cf.\ \cite{hb10,preethi14}), 
or from proper motions (comparing DSS with SDSS positions), or via a
detection in radio surveys. Our preliminary cross-correlation with object catalogues 
in these wavelength ranges leads us to estimate that at least 15\,\%
of the PHL and Ton objects are likely to be extragalactic.


\begin{acknowledgement}
HA \& ISB were supported by DAIP-UG grant \#318/13, and ISB by grants from CONACyT
and Universidad de Guanajuato (UG).
FRS \& WCC are grateful to Academia Mexicana de Ciencias (AMC) and UG
for summer research fellowships.~ Irina Andernach helped identifying objects on 
the DSS, and Roger Coziol made useful comments on the manuscript.
\end{acknowledgement}

\end{document}